# Positive and inverse isotope effect on superconductivity

T. D. Cao[*]

*Department of physics, Nanjing University of Information Science & Technology, Nanjing 210044, China*

**Abstract**

This article improves the BCS theory to include the inverse isotope effect on superconductivity. An affective model can be deduced from the model including electron-phonon interactions, and the phonon-induced attraction is simply and clearly explained on the electron Green's function. The focus of this work is on how the positive or inverse isotope effect occurs in superconductors.

## 1. Introduction

The phonon-mediated superconductivity is an important content of the BCS theory[1,2], and the BCS theory is supported with the isotope effect on common superconductors. Because various superconductivities are related to phonons except the superconductivities[3] near the Mott metal-insulator transition, the role of phonons is an important topic. Moreover, the isotope index may have either large or small value, and it may be positive or negative number[4,5,6], these features exceed the well-known BCS approximation. Could the inverse isotope effect be explained on phonons? Moreover, how should we evaluate the BCS approximation? This article will give a clear explanation to all these problems with Green's function.

---

[*]Corresponding author.
[*]E-mail address: tdcao@nuist.edu.cn (T. D. Cao).
[*]Tel: 011+86-13851628895

## 2. Green's function of normal state

To find the electron Green's function, we take this model

$$H = \sum_{k,\sigma} \xi_k c^+_{k\sigma} c_{k\sigma} + \sum_q \omega_q a^+_q a_q + \sum_{\substack{k,k',q \\ \sigma,\sigma'}} V_q c^+_{k+q\sigma} c_{k\sigma} c^+_{k'-q\sigma'} c_{k'\sigma'} + \sum_{k,q,\sigma} D_q c^+_{k+q\sigma} c_{k\sigma} A^{(+)}_q \quad (1)$$

where $V_q$ are the screened electron-electron interaction, $A^{(\pm)}_q = a_q \pm a^+_{-q}$, $c_{k\sigma}$ or $a_q$ destroys an electron or a phonon respectively, and $k \equiv \vec{k}$. The Green's function is defined in

$$G(k,\sigma,\tau-\tau') = -<T_\tau c_{k\sigma}(\tau) c^+_{k\sigma}(\tau')> \quad (2)$$

This function can be found by establishing the dynamic equation

$$\frac{\partial}{\partial \tau} G(k,\sigma,\tau-\tau') = -\delta(\tau-\tau')$$
$$+ \xi_k <T_\tau c_{k\sigma}(\tau) c^+_{k\sigma}(\tau)> + \sum_{q,\sigma} D_q <T_\tau c_{k-q\sigma} A^{(+)}_q c^+_{k\sigma}(\tau')> + \sum_{k',q,\sigma'} 2V_q <T_\tau c^+_{k'-q\sigma'} c_{k'\sigma'} c_{k-q\sigma} c^+_{k\sigma}(\tau')> \quad (3)$$

To solve this equation, we must calculate many-body functions such as $<T_\tau c_{k-q\sigma} A^{(\pm)}_q c^+_{k\sigma}(\tau')>$, and so on. If the effects of strong correlation are neglected, we get the final equation of normal state

$$[-i\omega_n + \xi_k] G(k,\sigma,i\omega_n)$$
$$= -1 + \sum_q 2V_q n_F(k-q\sigma) G(k\sigma,i\omega_n) - \sum_{q,\sigma} |D_q|^2 \{<A^{(+)}_{-q} A^{(+)}_q>(i\omega_n - \xi_{k-q}) + \omega_q <A^{(+)}_{-q} A^{(-)}_q> + 2\omega_q$$
$$\cdot [n_F(k-q,\sigma) + n_c \delta_{q,0}]\} [(i\omega_n - \xi_{k-q})^2 - \omega_q^2]^{-1} G(k\sigma,i\omega_n) \quad (4)$$

Because the strong correlation has been neglected, the $\Sigma(k,\sigma,i\omega_n)$ do not depend on the spin index, thus we rewrite the Green's function in

$$G(k,i\omega_n) = \frac{1}{i\omega_n - \xi_k - \Sigma(k,i\omega_n)} \quad (5)$$

where

$$\Sigma(k,i\omega_n) = -\sum_q 2V_q n_F(k-q) + \sum_q |D_q|^2 <A^{(+)}_{-q} A^{(+)}_q>(i\omega_n - \varepsilon_{k,q}) [(i\omega_n - \xi_{k-q})^2 - \omega_q^2]^{-1} \quad (6)$$

## 2. Phonon-induced attraction

It is shown that the electron-phonon interactions lead the imaginary part $\text{Im}\Sigma(k,\omega) \neq 0$, which affects the lifetimes of quasi-particles, and the real part of the self-energy to be

$$\mathrm{Re}\Sigma(k,\omega) = -\sum_q 2V_q n_F(k-q) + \sum_q |D_q|^2 <A^{(+)}_{-q}A^{(+)}_q>(\omega-\varepsilon_{k,q})\,[(\omega-\xi_{k-q})^2-\omega_q^2]^{-1} \qquad (7)$$

The excitation energies $E_k$ in normal state are determined by

$$E_k - \xi_k - \mathrm{Re}\Sigma(k,E_k) = 0 \qquad (8)$$

After comparing with $\mathrm{Re}\Sigma(k,\omega) = \sum_q 2V_q^{eff} n_F(k-q)$, we find that phonons provide an attraction between electrons when the excitation energies meet

$$(E_k - \varepsilon_{k,q})\,[(E_k - \xi_{k-q})^2 - \omega_q^2]^{-1} > 0 \qquad (9)$$

and

$$\varepsilon_{k,q} = \xi_{k-q} - <A^{(+)}_{-q}A^{(+)}_q>^{-1}\{\omega_q <A^{(+)}_{-q}A^{(-)}_q> + 2\omega_q n_F(k-q)\} \qquad (10)$$

Where $n_c \delta_{q,0}$ is neglected since the phonon of $q=0$ is neglected. Concretely speaking, phonons may induce the attraction when the excitation energies $E_k$ in the second term of Eq.(7) meet the condition $\xi_{k-q} - \omega_q < E_k < \xi_{k-q} + \omega_q$ for $E_k < \varepsilon_{k,q}$. The condition represents the requirement of a negative part of $V_q^{eff}$ for the excitation energies with wave vector $\vec{k}$. Therefore, the possibility of phonons inducing attraction increases with the phonon frequency. This condition can be written in $E_k = \xi_{k-q} + \gamma\omega_q$ ($-1 < \gamma < 1$). Use $E_k = \xi_{k-q} + \gamma\omega_q$, $<A^{(+)}_{-q}A^{(-)}_q> = -1$ and $<A^{(+)}_{-q}A^{(+)}_q> = 1 + 2n_B(q)$, Eq.(8) becomes

$$E_k = \xi_k - \sum_q 2V_q n_F(k-q) + \sum_q |D_q|^2 \frac{\gamma + [1+2n_F(k-q)]/[1+2n_B(q)]}{(\gamma^2-1)\omega_q} \qquad (11)$$

Note $n_B(q) \equiv n_B(\omega_q)$ for the phonon number with frequency $\omega_q$, the third term of Eq.(11) can be positive, thus phonons provide an attractive part for the affective interaction $V_q^{eff}$ as shown in Eq.(12). Some materials are not superconductors could be understood on Eq.(11). Besides other pair-breaking effects, some materials are not superconductors because $|D_q|$ is not large enough and $V_q$ not small enough. Au, Ag, Cu etc. belong such examples.

Let us consider the mass dependence of parameters. Since $|D_q|^2 \propto 1/M\omega_q$ and $\omega_q \propto M^{-1/2}$ for the monatomic lattice, thus $|D_q|^2/\omega_q$ do not depend on the mass of atom. When Eq.(1) includes the optical phonons, problems will become complex. If each cell contains $r$ atoms, we find $|D_q|^2/\omega_q \to \sum_{sj} |D_q^{js}|^2/\omega_q^j \to \propto \sum_{sj} \lambda_s/M_s(\omega_q^j)^2$ ($s=1, ..., r$; $j=1, ..., 3r$), where $j$ designates a branch. For simplicity, consider only a crystal with two atoms per unit cell. We have the mass dependence $(\omega_q^j)^2 \sim 1/M_1$ for the optical branch near the surface of the Brillouin zone, and $|D_q|^2/\omega_q \to \lambda_1 + \lambda_2 M_1/M_2$, thus $|D_q|^2/\omega_q$ increases with $M_1$ while decreases with $M_2$. Therefore, the attraction could increase with the mass of element "1", which may be related to the inverse isotope effect. When the attraction decreases with the mass of element "2", the isotope effect may occur.

However, whether the affective interaction $V_q^{eff}$ is attractive will also depend on the screened interaction $V_q$. The affective interactions are included in the model

$$H^{eff} = \sum_{k,\sigma} \xi_k c_{k\sigma}^+ c_{k\sigma} + \sum_{\substack{k,k',q \\ \sigma,\sigma'}} V_q^{eff} c_{k+q\sigma}^+ c_{k\sigma} c_{k'-q\sigma'}^+ c_{k'\sigma'} \qquad (12)$$

It leads to this BCS gap equation

$$\Delta(k) = \sum_q V_q^{eff} \Delta(k-q) \frac{n_F(E_{k-q}) - n_F(-E_{k-q})}{2E_{k-q}} \qquad (13)$$

where $E_k = \sqrt{\tilde{\xi}_k^2 + \Delta^2(k)}$ and $\tilde{\xi}_k = \xi_k - \sum_q 2V_q^{eff} n_F(k-q)$. This equation shows that only the excitation energies near Fermi surface contribute to the gap, thus $\vec{k} - \vec{q} \sim \vec{k}_F$ in Eq. (13). Again Eq.(7) shows that $V^{eff}(q)$ may be negative for $\vec{k} - \vec{q} \sim \vec{k}_F$ with $E_k = \xi_{k-q} + \gamma\omega_q$ ($-1 < \gamma < 1$), and this leads to $T_c \neq 0$ in Eq.(13). However, both textbooks and literatures usually suggest that "the phonons provide an attractive interaction between electrons for $\omega < \omega_q$ ($\mu=0$)", and this seems that the attraction would appear among the electrons in low energy states. In contrast with this suggestion, our results may accord with physics, and we provide BCS theory more reasons: BCS theory take $\Delta$ as a constant number, because they take $k - q \sim k_F$ ($= \vec{k}_F$) and $k \sim k_F'$ ($= \vec{k}_F'$) to meet $\Delta_{k-q \sim \vec{k}_F} = \Delta_{k \sim \vec{k}_F'}$ =constant for the s-wave symmetry, thus $V^{eff}(q)$ are taken as $-V$ in the thin sphere-shell around the Fermi surface ($E_{k-q} \sim 0$ and $T \to T_c$ in Eq.(13)). These approximations lead to the well-known result

$$k_B T_c = 1.14 \omega_c e^{-1/VN_F} \qquad (14)$$

The condition $\xi_{k-q} - \omega_q < E_k < \xi_{k-q} + \omega_q$ shows that the cut-off frequency $\omega_c$ (the BCS theory takes it as $\omega_D$) increases with $\omega_q$, thus the mass dependence of $\omega_c$ is determined by $\omega_q$, and $\omega_c$ should decrease with the masses of

## 4. Positive and inverse isotope effect

Our discussion above arrives at these conclusions:

(ⅰ) The cut-off frequency $\omega_c$ usually decreases with the masses of the atoms, while $\omega_c$ weakly depend on the masses of the atoms if the optical phonons dominate the superconductivity.

(ⅱ) The affective attraction $V$ could either increase or decrease with the mass of some atom, which are determined by the mass dependence of the phonon frequency.

The positive and inverse isotope effect can be explained with these conclusions. Eq.(13) shows that the calculation of the superconducting transition temperature can be done by the cut-off frequency $\omega_c$ and the affective interaction $-V$ ($\omega_c$ is taken as $\omega_D$ in BCS approximation), and $T_c$ increases with both $\omega_c$ and $V$ as shown in Eq.(13). On the basis of the two conclusions above, when the affective attraction $V$ decreases with the mass of some atom, the inverse isotope effect may occur, while other cases favor the (common) isotope effect.

Although Eq.(8) shows that the phonons affect the electronic structure, and Eq.(13) shows that the electronic structure affect the superconducting transition temperature, by the way, these influences from the mass dependence of the electronic structure should be neglected for the superconductivity.

In summary, the BCS approximation is improved in this article. It is shown that the phonon-induced attraction can be understood on the Green's function of normal state, and the positive and inverse isotope effect can be explained with phonons.

**Acknowledgments**

I thank Nanjing University of Information Science & Technology for financial support.